\documentclass[12pt]{article}
\usepackage{amsmath}
\usepackage{amssymb}
\usepackage{epsfig}
\usepackage{cite}
\usepackage{color,colordvi}

\begin{document}
\vspace{0.01cm}
\begin{center}
{\Large\bf   Quantum  Exclusion  of Positive Cosmological Constant?} 

\end{center}

\vspace{0.1cm}

\begin{center}

{\bf Gia Dvali}$^{a,b,c}$ and {\bf Cesar Gomez}$^{b,e}$\footnote{cesar.gomez@uam.es}

\vspace{.6truecm}


{\em $^a$Arnold Sommerfeld Center for Theoretical Physics\\
Department f\"ur Physik, Ludwig-Maximilians-Universit\"at M\"unchen\\
Theresienstr.~37, 80333 M\"unchen, Germany}


{\em $^b$Max-Planck-Institut f\"ur Physik\\
F\"ohringer Ring 6, 80805 M\"unchen, Germany}

{\em $^c$Center for Cosmology and Particle Physics\\
Department of Physics, New York University, \\ 
4 Washington Place, New York, NY 10003, USA}

{\em $^e$
Instituto de F\'{\i}sica Te\'orica UAM-CSIC\\
Universidad Aut\'onoma de Madrid,
Cantoblanco, 28049 Madrid, Spain}\\

\end{center}


\begin{abstract}
\noindent  
 
{\small

 We show that a positive cosmological constant is incompatible with the quantum-corpuscular  resolution 
 of de Sitter metric in form of a coherent state.   
 The reason is very general and is due to the quantum self-destruction of the coherent state because of the  scattering of constituent graviton quanta.    
 This process creates an irreversible quantum clock,  which precludes eternal de Sitter. 
 It also eliminates the possibility of Boltzmann brains and Poincare recurrences.   
 This effect is expected to be part of any microscopic theory that  takes into account  the quantum corpuscular structure of the cosmological background.  This observation puts  the cosmological constant problem in a very different light, promoting it,  from a naturalness problem, into a question of quantum consistency. We are learning that quantum gravity cannot tolerate exceedingly-classical sources.}

\end{abstract}

\thispagestyle{empty}
\clearpage

\section{Changing the Question}

  The failure of dynamically solving the cosmological constant problem for so many years, should push us  to reconsider  the  rightness  of the question we are asking.  Instead of asking: 
  
  $~~~$
  
  {\it  Why is the cosmological constant so unnaturally small?}  
  
  $~~~$
   
     We should be asking:
  
  $~~~$
     
    {\it  Is the cosmological constant a quantum-mechanically-consistent notion?}  
  
  $~~~$
      
   In this paper, following \cite{us},  we shall make a little step in justifying 
    this line of thought.   We shall uncover that  a positive constant source is incompatible with the quantum-corpuscular resolution of gravity.   Our main results were already given
    in \cite{us},  but in this note we shall try to present the discussion in a more convenient and focused  form. 
  
\section{Quantum-Corpuscular View}

 The classical  field configurations are limits of quantum states with large occupation 
 numbers of particles.  For example,  a classical radio-wave is in reality a coherent state of photons.  Of course, in the quantum world, the number-density of finite-wave-length photons in a finite-energy-density wave,  is finite.   This finiteness must be taken into account in the quantum treatment.    
 
 Surprisingly, this obvious corpuscular view  has not been extended to classical gravitational backgrounds, such as black holes and cosmological spaces,  until very recently. 
 In \cite{usBH},\cite{us} we have developed such a corpuscular treatment.  Consider  
 a classical gravitational field of a characteristic curvature radius $R$, much larger than the Planck length, $L_P$.   In our corpuscular picture, this field must be treated as a composite entity,  a sort of coherent state or a condensate,  of gravitational quanta of mean wavelength $R$.  The fact that the wave-length is determined by the classical radius of the problem, as opposed to some shorter 
 UV-scale, is crucial for the reliable estimates of the interactions among them.   
 
  It then immediately follows  that the occupation number is given by,
   \begin{equation}
 N \, =\, {R^2\over L_P^2} \, .
 \label{N}
 \end{equation}
    
     For black holes, the correctness of such an approach can be justified by the following thought experiment.  Let us attempt to form a black hole of radius $R$, by putting together gravitons 
     of this wavelength. Since $R \, \gg \, L_P$, the quantum graviton-graviton coupling, 
    $\alpha \equiv L_P^2 /R^2$, is very weak.  Thus, individual graviton pairs attract very weakly.  In such a situation,  gravitons can only be kept together by a collective binding potential.  Its strength can be   parametrized by the collective coupling $N\alpha$.   As long as this coupling is not strong, the mass of the system can be estimated by summing up the energies of individual gravitons,  
     \begin{equation}
 M \, =\, N { \hbar \over  R} \, . 
 \label{M}
 \end{equation}
 Now, requiring that the gravitational radius of this system,  $MG_N$, is equal to $R$, we 
 arrive to (\ref{N}).  Several things are remarkable about this number. 
 First, it is equal to $1/\alpha$.   Thus, the  system becomes  a black hole, exactly when the 
 collective quantum effect of gravitons becomes strong enough for binding them together.    
  Such a system of gravitons is at the {\it quantum critical point}  separating the phases of free gravitons and  of a bound-state.  Another immediate observation is that the number  happens to coincide with the Bekenstein-Hawking entropy, although 
the entropy was never mentioned to start with.    As it was suggested in \cite{usBH},  this equality is not a coincidence and entropy and other black hole properties indeed emerge from quantum criticality of graviton condensate.      
           
            We shall not consider black holes in this framework further.  Instead, we shall turn to its cosmological implications. 
   The corpuscular treatment was applied to cosmology in \cite{usBH, us}.    In \cite{us},  it was discovered that the quantum corpuscular effects are in conflict with the notion of eternal de Sitter.    
  In this short note, we wish to focus solely on this aspect,  which has far-reaching consequences.  This picture provides an explicit  microscopic framework in which  the question of the quantum consistency of the cosmological constant can be posed.   
      
       \section{Quantum-Corpuscular View of  de Sitter}  
          
           In order to get straight to the point, let us consider a de Sitter space.
     Classically, such a space is sourced by a strictly-constant vacuum energy density $\Lambda$.                 
    It determines  the Hubble parameter from Friedmann equation, 
       \begin{equation}    
       H^2 \,   =  \, {8\pi \over 3}  {\Lambda \over M_P^2} \,.  
       \label{Hclass}
       \end{equation}
         Thus, classically the system is described by a single quantity, $H$  that is determined by the constant source  $\Lambda$
    via GR equations.  Since $\Lambda$ is constant, there is no  classical clock.  Thus, in this picture  de Sitter space  described in terms of  $H$ and $\Lambda$ can be {\it eternal in future}.         
         
    We want to show, that the above  is no longer true in the quantum-corpuscular theory.  Here, the classical field dynamics  is replaced by a mean-field description of some multi-particle state. 
       Our purpose is to show, that even if the mean-field  description, in terms of $H$ satisfying  classical GR, exists at some 
  fixed time $t=t_0$,  it will cease to exist after the time, 
  \begin{equation}
   \Delta t_{max} \, \sim  \,  { M_P^{5} \over \Lambda^{3/2}} \,.     
   \label{timemax}  
  \end{equation} 
 In other words,  (\ref{timemax}) is the {\it maximal service time}  of the classical de Sitter. 
%
%
 
    We shall try to keep the argument very general 
    and maximally insensitive  to  the particular details of the quantum resolution.  We thus assume 
    that in the quantum theory the de Sitter  Hubble patch has to be replaced by a multi-particle coherent state 
    of constituent gravitons.  We shall denote it by 
    $|N \rangle$.  Here $N$ is a  properly-normalized expectation value of the number operator
   of the constituent particles,  $N \equiv \langle N| \hat{N} |N \rangle$.   The multi-particle state 
  $|N \rangle$ is defined in such a way that in mean-field approximation reproduces the dynamics of the classical 
  gravitational field of GR.

    
      The precise nature of the constituent gravitons is unimportant for our argument, as along as they exist and satisfy some very general property. Namely, the dominant frequency of the constituents is set 
    by the Hubble constant $H$.  In such a case, their re-scattering  rate can be estimated reliably.   
    
       Of course, 
   this state is a distribution   
   $|N \rangle \,  = \,\prod_{k}  \, |N(k)\rangle\, $ over the coherent states of different frequency $k$, $|N(k)\rangle \, = {\rm e}^{-{N(k) \over 2}} \, \sum_0^{\infty} \, {N(k)^{{n_{k}\over 2}} \over \sqrt{n_{k}!}}  |n_{k}\rangle \,$.  However, by assumption the dominant $k$ is set by $H$. 
   
   Let us assume that at some $t=t_0$ the mean-field 
   description of the state $|N \rangle$ is well approximated by classical GR and let us see 
   how long it will take for the system to depart from this description. 
     Initially, while GR is a good mean-field picture, according to (\ref{N}), the graviton occupation number is 
     \begin{equation}
     N \, =  \, {M_P^2 \over H^2}\,  =  \,   {M_P^4 \over \Lambda}.   
   \label{Ngr}
   \end{equation}
   In particular,  GR is a valid mean-field 
  description as long as the expectation value of $\hat{N}$ is related to  $\Lambda$ via the equation (\ref{Ngr}). 
   We now come to our most important point:  
   
   $~~~$
   
  {\it  1) There inevitably exist quantum transitions, which violate 
   the coherence of the state;

   2) The rate of these transitions is  $H$.  } 
   
   $~~~$
     
    As a result of these coherence-violating transitions the expectation values no longer satisfy the
   relations of classical GR. Thus, we uncover a quantum clock which advances at Hubble rate. This quantum clock  abolishes the possibility of  
   eternal de Sitter. 
   
    For simplicity of estimate, assume that the initial coherent state  $|N \rangle$,
 which satisfies GR constraints, 
   consists solely of $k = H$-frequency gravitons. Due to interactions between constituent gravitons, such a state 
 experiences transitions into non-coherent states. 
   For example,  a transition during which $N$  decreases by two units and a graviton-graviton pair of energies
$k, k' \sim H$,  is produced.    The new state is  $|N-2,1_k, 1_{k'} \rangle$, where 
only the $N-2$ part is coherent. 
 This process takes place in first order in $\alpha$,  but is enhanced due to large 
   occupation number of gravitons. The corresponding  matrix element is, 
   \begin{equation}
   |\langle N-2, 1_k, 1_{k'}|a^+a^+ a a |N \rangle|^2 \, \sim  \alpha^2 N^2 \, \sim \, 1 \, . 
  \label{matrix} 
   \end{equation}  
  Here  $ a$  and  $a^+$ are the annihilation and creation operators of initial and final 
  particles respectively.  The graviton-graviton coupling corresponding to the characteristic momentum exchange 
  in this process is,  $\alpha\, \sim  \, H^2/M_P^2 \, \sim N^{-1}$.

      The physical meaning of the above matrix element is very transparent. 
     The  transition process is due to re-scattering of the background gravitons, 
      \begin{equation}
        graviton + graviton \rightarrow \,  graviton + graviton \, .
       \label{process} 
       \end{equation}       
  In this process the final particles are not in the coherent state.  
   Thus, they are no longer a part of the classical background. 
    Correspondingly, the occupation number of constituent gravitons in the coherent state decreases. 
   The rate of this leakage is given by 
  \begin{equation}
   \Gamma \, = \, H \, \alpha^2N^2 \, = \,  H \, .
    \label{rate} 
    \end{equation}
  The factor $H$ comes from the characteristic momentum-transfer in the process. 
 The other factors are the same as in the matrix element (\ref{matrix}) \footnote{
 This process of particle-creation can be viewed as the quantum origin of Gibbons-Hawking radiation in our picture. 
  The basic assumption of our corpuscular approach differs dramatically from the standard view, where a Hubble patch is quantum-mechanically associated with a thermal density matrix $\rho$ of temperature equal to the Gibbons-Hawking temperature and with entropy $S= - tr(\rho \ln \rho)$ equal to Gibbons-Hawking de Sitter entropy \cite{gibbons}.  Nevertheless,  we recover the Gibbons-Hawking thermal radiation as the result 
 of depletion of a pure coherent state in  the limit $N = \infty$. In this limit the back reaction can be ignored and 
 the coherent state can model an eternal classical field.}.

%
%
%
%

    Thus, the {\it quantum}  change of the occupation number of gravitons 
per Hubble patch is given by the following time-evolution equation,  
\begin{equation}
\dot{N} \, \sim  \, - \, H\, .
\label{quantum!}
\end{equation}
Notice,  that this variation counts the number of particles, that have abandoned the coherent state. 
Thus, after every Hubble time,  the final state is slightly less coherent that the initial one and GR description becomes less accurate.  The process continues irreversibly until the departure from coherence becomes order one, at which point 
one can no longer talk about any sensible classical de Sitter space.

 Equation (\ref{quantum!}) makes the impossibility of eternal de Sitter clear. Indeed, the necessary  
condition for such eternity is that  $H$ is constant. 
Then, according to classical GR,  so must be $N$.    However,  $N$ continues to evolve 
in a non-coherent way due to the quantum depletion. That is, within each Hubble time, roughly  
one graviton leaves the coherent state. 
   Thus, the system 
will go out of validity of GR within a number of Hubble times of order  
$N$. 

 Thus, even if we stop the classical clock, the quantum clock continues to advance and 
finally puts the system out of the classical regime. 

Notice, that the negative sign in front of  the quantum-depletion term in  (\ref{quantum!}) should not make us to think that leakage can be interpreted as the increase of Hubble.  This is not the case.   Leakage 
is an intrinsically-quantum process that leads to the violation of coherence and departure 
from GR-description.  So, it cannot be parametrized by the variation of classical GR quantities, such as
$H$.

    Notice, that in our reasoning we do not hit any strong  coupling point. Our estimates of depletion rate continues to be reliable even after the classical description breaks down and one can no longer talk about
 classical de Sitter type metric.

   Requiring that throughout its evolution the mean field description of the system stays within the range of validity of classical GR  
we get the bound on maximal duration of classical de Sitter (\ref{timemax}). 
 This bound describes the time during which the Hubble patch evolves into an intrinsically-quantum state for which the mean-field description is, not even remotely,  related to the classical GR.  
  At this pint, the Universe, no matter how big, becomes quantum. 
  
   The fact that quantum effects can significantly 
  correct mean-field evolution over long time scales is not so unusual. For example,  the proton decay 
 in grand unified theories is extremely weak.  However,  over a period of $10^{34}$ years (a typical life-time of proton in GUTs) it dramatically corrects the classical evolution of the macroscopic water-tank.      

\section{Quantum Exclusion of Boltzmann Brains and of Poincare Recurrences}

  Within the semi-classical treatment, it has been known for a long time that there exist 
  exponentially-rare quantum fluctuations  that can violate de Sitter invariance.   Due to this fluctuations, various non-perturbative objects can be produced. A good example is the creation of topological defects in de Sitter space \cite{vilenkin}.  Even if the theory contains no other massive object,  one can always violate de Sitter invariance by producing black holes.  
     The rates for such processes are exponentially-suppressed by a factor $\sim e^{-S_E}$, where 
     $S_E$ is the Euclidean action of an instanton responsible for the production of a given object.  For an object of mass 
   $M$,  typically we have $ S_E \sim M/H$,  which also reproduces a Boltzmann suppression factor for a thermal bath of Gibbons-Hawking temperature $T_{GH} \sim H$.   Correspondingly, the time-scale  required for materializing such fluctuations is {\it exponentially long},
   \begin{equation}
    \Delta t_{Boltz} \, \sim \, H^{-1} e^{S_E}\, .
   \label{thermal} 
  \end{equation}  
 We shall refer to this time as the Boltzmann time.
 
  As a particular example, the action $S_E$ can belong to a Boltzmann brain, a self-aware entity created by quantum fluctuations.
   Hence, in the standard picture of eternal de Sitter, one only has to wait long enough  in order to 
   witness creation of Boltzmann brains and various other objects. 
         
  An even longer time-scale is the Poincare recurrence time \cite{Susskind}.  Such recurrence can take place if one assumes that the 
de Sitter Hilbert space is finite \cite{Banks} \footnote{Strictly speaking, the only necessary requirement is to have a hamiltonian with countable spectrum as part of the de Sitter symmetry group \cite{Susskind}.}.  The Poincare recurrence time is then given by 
$\Delta t_{Poincare} \sim H^{-1} e^{N}$. Where, $N$ is the  Gibbons-Hawking entropy of de Sitter, which coincides with the occupation number of coherent-state gravitons in our picture. 

    An important point that the corpuscular picture brings across is that all the above type processes 
    essentially have no chance of  being materialized, because the de Sitter service time (\ref{timemax}) is much shorter then the Boltzmann and Poincare recurrence times. Thus, the corpuscular quantum clock renders the exponentially-rare quantum processes essentially irrelevant.  
    
    We can find out which entities have a chance to materialize within the de Sitter service time by 
    equating the two time scales (\ref{timemax}) and (\ref{thermal}). This implies that the only entities that can materialize within the de Sitter service time  are the ones with Euclidean action satisfying the upper bound.     
    \begin{equation}
          S_E \, = \, ln (N) \, .
    \end{equation}
    However, even such low-action objects will never be produced within the de Sitter service time, if their mass is well above the  
    Hubble.  
   This excludes the Poincare recurrence, as well as any sensible Boltzmann brain creation \footnote{Exactly the same phenomena precludes the existence of eternal inflation (see \cite{us}).}. 
   
   The latter statement requires some clarification.   For this, let us estimate the rate of 
   Boltzmann brane creation in a de Sitter space of some fixed Hubble $H$.  Conservatively, we may assume that the Boltzmann brain has a memory-capacity approximately that of a human 
   brain.    We shall take this to be abut $10^{12-13}$ bits.  The most compact  object with such a capacity of information-storage is 
   a black hole of mass  $M \sim 10^6 M_P$.   Then, the Boltzmann time required for  producing such an object is  
     $ \Delta t_{Boltz} \, \sim \, H^{-1} e^{(10^6M_P/H)}$.  It is obvious that this time   
  exceeds the de Sitter service time (\ref{timemax}) by many orders  of magnitude, for arbitrary value of Hubble.

  \section{Conclusions} 
  
  Our findings have obvious implications for the cosmological constant problem.  The puzzle that is usually posed 
  as a problem of severe fine tuning, now appears in a completely different light and becomes 
  a question of microscopic quantum consistency.  
           What we are learning is that the corpuscular quantum theory cannot tolerate eternal 
           classical sources.
 The reason is an inevitable conflict between the  classical mean-field description and quantum evolution. 
 
   For constant $\Lambda$ there is no classical clock. However, there still exists a quantum one, which comes from the quantum depletion  of constituent gravitational corpuscles. 
       This is due to graviton-graviton scattering, with the rate given by  (\ref{rate}).         
       This quantum clock represents an underlying corpuscular mechanism for Gibbons-Hawking radiation \cite{gibbons} . 
      Except,  the corpuscular picture reveals its hidden side. What in semi-classical theory is seen as particle-creation out of the vacuum, in reality is the leakage of real corpuscles of the coherent state. Due to this leakage, 
      any classically-sensible de Sitter state can last maximum $N$ Hubble times.   Thus, the maximal duration of 
      de Sitter is  (\ref{timemax}). After this time, de Sitter will evolve into a state 
      $|quant\rangle$, which has no mean-field description in terms of classical GR.  The
   solution of the cosmological constant problem hangs upon the quantum consistency of the state
   $|quant\rangle$.  Answering this question in beyond the scope  of the present paper. However, regardless 
   what is the answer, the quantum corpuscular view fundamentally changes the nature of the question\footnote{In our picture,  Gibbons-Hawking temperature of de Sitter can be understood in parallel with the Hawking temperature of a black hole.  Thus, in both cases,  due to the decay of the coherent state, the  mean-field description can diverge from classical GR after the time $t \sim N^{3/2} /M_P$.  However, there is an important difference between the two cases. 
 In sharp contrast from de Sitter,  such departure from classical GR is not an inconsistency for a black hole.  This is because for a black hole there is no external constant classical source in form of $\Lambda$.}.

   
    We must stress that  we do not claim that quantum mechanical back reaction renormalizes or in some other way destabilizes $\Lambda$, as this is the case in semi-classical studies \cite{infra1,infra2}. We 
 sympathize with these approaches, but  our claim is very different.  Our findings indicate that constant $\Lambda$ is  incompatible 
    with the quantum corpuscular view coming from finiteness of $N$.    The effects we are uncovering 
    disappear  in semi-classical limit, in which $N \rightarrow \infty$.  In this limit, classical de Sitter becomes eternal.  
    
    We are learning that the quantum corpuscular resolution of de Sitter geometry, with finite $N$, cannot tolerate exceedingly classical sources.   The reason is the secret conflict between the classicality of the source and 
    quantum-corpuscular structure of the gravity it sources. 
   The cosmological  constant being the only known truly-classical source appears to be incompatible 
  with a quantum understanding of $N$.

\section*{Acknowledgements}
 
The work of G.D. was supported by Humboldt Foundation under Alexander von Humboldt Professorship,  by European Commission  under ERC Advanced Grant 339169 ``Selfcompletion'',   by TRR 33 \textquotedblleft The Dark
Universe\textquotedblright\   and  by the NSF grant PHY-0758032. 
The work of C.G. was supported in part by Humboldt Foundation and by Grants: FPA 2009-07908, CPAN (CSD2007-00042) and by the ERC Advanced Grant 339169 ``Selfcompletion'' .

\end{document}